\documentclass[pra,aps,reprint,citeautoscript,amsmath,amssymb,superscriptaddress]{revtex4-1}
\usepackage{graphicx}
\usepackage{dcolumn}
\usepackage{bm}
\usepackage{amsmath}
\usepackage{hyperref} 
\usepackage{soul} 
\DeclareGraphicsExtensions{.pdf,.eps,.png,.jpg,.mps}
\usepackage{mathrsfs}
\usepackage{soul} 
\usepackage[utf8]{inputenc} 
\usepackage{hyperref} 
\usepackage{xcolor} 
\usepackage{amsmath}
\usepackage{amsthm}
\usepackage{eucal}
\usepackage{setspace}

\usepackage{booktabs}

\begin{document}
\title{Proximity-induced spin-polarized magnetocaloric effect in transition metal dichalcogenides}

\author{Natalia Cortés} 
\email{natalia.cortesm@usm.cl}
\affiliation{Departamento de Física, Universidad Técnica Federico Santa María, Casilla 110V, Valparaíso, Chile}
\author{Francisco J. Peña}
\affiliation{Departamento de Física, Universidad Técnica Federico Santa María, Casilla 110V, Valparaíso, Chile}
\author{Oscar Negrete}
\affiliation{Departamento de Física, Universidad Técnica Federico Santa María, Casilla 110V, Valparaíso, Chile}
\affiliation{Centro para el Desarrollo de la Nanociencia y la Nanotecnolog\'ia, 8320000 Santiago, Chile}
\author{Patricio Vargas}
\affiliation{Departamento de Física, Universidad Técnica Federico Santa María, Casilla 110V, Valparaíso, Chile}
\affiliation{Centro para el Desarrollo de la Nanociencia y la Nanotecnolog\'ia, 8320000 Santiago, Chile}

\date{\today}


\begin{abstract}
We explore proximity-induced magnetocaloric effect (MCE) on transition metal dichalcogenides, focusing on a two-dimensional (2D) MoTe$_2$ monolayer deposited on a ferromagnetic semiconductor EuO substrate connected to a heat source. We model this heterostructure using a tight-binding model, incorporating exchange and Rashba fields induced by proximity to EuO, and including temperature through Fermi statistics. The MCE is induced on the 2D MoTe$_2$ layer due to the EuO substrate, revealing large spin-polarized entropy changes for energies out of the band gap of the MoTe$_2$-EuO system. By gating the chemical potential, the MCE can be tuned to produce heating for spin up and cooling for spin down across the $K$ and $K'$ valley splitting in the valence band, whereas either heats or cools for both spins in the conduction band. The Rashba field enhances the MCE in the valence zone while decreasing it in the conduction bands. The exchange field-induced MCE could be useful to produce tunable spin-polarized thermal responses in magnetic proximitized 2D materials.
\end{abstract}


\maketitle
\date{Today} 
\textit{Introduction}. The magnetic proximity effect takes place when the magnetization of a magnetic crystal is induced on a neighboring nonmagnetic material \cite{vzutic2019proximitized}. Diverse nonmagnetic materials are used to react to the induced magnetism, but considerable attention is lately given to atomically thin two-dimensional (2D) layers \cite{geim2013van,yazyev2015mos2,ajayan2016van}. The thinness of 2D layers allows for short-range induced magnetism, driving to modifications in the combined electronic states \cite{vzutic2019proximitized}, as seen in the band structure responses. One attractive class of 2D materials that can be mixed with magnetic crystals to produce magnetic proximity effects include transition metal dichalcogenides (TMDs) of the semiconducting $MX_{2}$ family ($M=$ Mo and W; $X=$ S, Se and Te) \cite{liu2015electronic}. The intrinsic lack of inversion symmetry and spin-orbit coupling (SOC) in TMDs cause a sizable spin splitting at the valence band edges of the spin-valley coupled $K$ and $K'$ ---degenerated, yet inequivalent--- valleys in the Brillouin zone (BZ) \cite{manzeli20172d}, which are related each other by time-reversal symmetry (TRS) \cite{Xiao2012}. 

Valley splitting to encodes information requires lifting the degeneracy of $K$ and $K'$ TMD valleys, which can be achieved due to broken TRS by either, an external magnetic field \cite{aivazian2015magnetic}, or induced magnetic exchange fields (MEFs) driven by ferromagnetic substrates \cite{Zhao2017,Zhong2017van,Seyler2018,zou2018probing}; however, the resulting valley splittings because of the former are small ($\simeq 0.1-0.2$ meV/T) \cite{li2014valley,macneill2015breaking}. The induced MEFs on 2D TMDs may have some advantages over large external magnetic fields needed to break TRS and achieve valley polarization. For example, a giant valley splitting (300 meV) at zero Kelvin was predicted for MoTe$_2$-EuO \cite{qi2015giant}, a large valley splitting of 16 meV/T at a temperature of 7 Kelvin has been experimentally obtained due to the induced MEF in a WS$_2$-EuS heterostructure \cite{norden2019giant}, and a valley splitting of $\simeq 3.5$ meV was measured in WSe$_2$-CrI$_3$ at 5 K. As one can note, the thermal conditions of typical van der Waals experiments clearly play an important role, as the valley splittings of the proximitized TMDs show large dependence on the magnetic substrates Curie temperature \cite{Zhong2017van}.

Entropy is a useful fundamental thermodynamic quantity intimately related to temperature and accounting for the number of accessible states of a system. When a material is at constant temperature, and subjected to external magnetic field changes, it experiences entropy changes, and the magnetocaloric effect (MCE) arises cooling or heating the sample \cite{gomez2013magnetocaloric,miller2014magnetocaloric,franco2018magnetocaloric}. The MCE has been analyzed in diverse magnetic structures \cite{von2009understanding}, including a spin-gapped material \cite{chakraborty2019magnetocaloric}, a one-dimensional spin-1/2 system \cite{zhitomirsky2004magnetocaloric}, frustrated magnets \cite{zhitomirsky2003enhanced}, spin-1/2 2D lattices \cite{honecker2006magnetocaloric}, nanomagnets \cite{skomski2008temperature}, and superlattices showing large entropy changes due to exchange interactions \cite{mukherjee2009magnetocaloric}. In 2D layers of graphene and gold, the MCE shows an oscillatory behavior \cite{reis2013oscillating,reis2012oscillating2}, and an external magnetic field is capable to control the entropy in TMDs \cite{diffo2021thermodynamic}. The MCE has also been reported in bulk EuS \cite{hashimoto1981magnetic} and EuO \cite{ahn2005preparation}, as well as in EuO thin films \cite{lampen2021table}, finding maxima entropy changes near the Curie temperature of each ferromagnet. 

We propose here a novel MCE associated with changes of the induced EuO MEF on MoTe$_2$, driving to large entropy changes beyond the Curie temperature of EuO \cite{averyanov2018high}. The strength of the induced MEF could be modulated through van der Waals engineering of proximitized materials \cite{Zhong2017van} via nonmagnetic spacer layers \cite{Zhao2017}, or through biaxial strain \cite{li2018large}. Time-reversal symmetry breaking on MoTe$_2$ due to the MEF causes spin-polarized entropy production \cite{maes2003time,andrieux2007entropy}, then a spin-polarized MCE can be generated in the valley splitting energy zones. The MoTe$_{2}$-EuO heterostructure is modeled by a three-orbital tight-binding model (3OTB), and the MCE is derived from Fermi statistic. We find that when the Fermi energy is tuned along the $K$ and $K'$ valleys of the valence band, the MoTe$_{2}$-EuO system is able to heat for spin up and cools for spin down, while either heats or cools for both spins in the conduction band. We also analyze the effect of the Rashba field, showing that it enhances the cooling effects in the valence band. The generic existence of spin-polarized TMD electronic states given by TRS breaking that can be accessed by gating, suggests that these hybrid systems could be used as tunable thermal spin filters \cite{gholami2021pure} for functional applications.

\begin{figure}
\centering
\includegraphics[width=\linewidth]{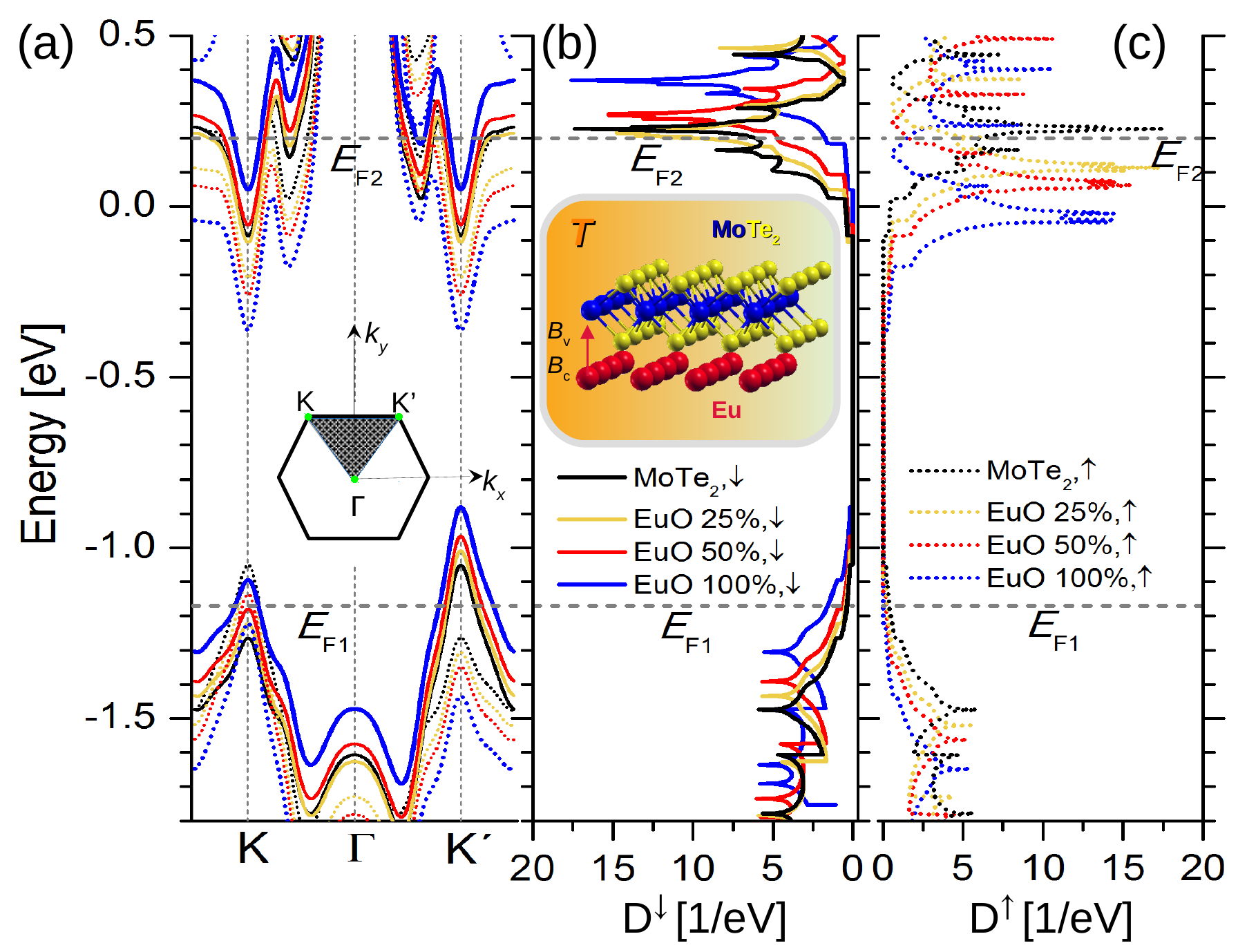}
\caption{(a) $K-\Gamma-K'$ band structure for suspended MoTe$_{2}$ monolayer (black lines) and MoTe$_{2}$-EuO with $\lambda_{R}=0$ (yellow 25\%, red 50\% and blue 100\% of $B_v$ and $B_c$). $\Gamma=(0,0)$, $K=\frac{2\pi}{3a}\big(-1,\sqrt{3}\big)$ and $K'=\frac{2\pi}{3a}\big(1,\sqrt{3}\big)$, with $a=3.56$ \AA\ the 2D MoTe$_2$ lattice constant. Spin-polarized DOS (b) $D^{\downarrow}$, (c) $D^{\uparrow}$. The inset in (a) shows the BZ for MoTe$_{2}$-EuO, the shaded area indicates where the DOS is calculated in (b) and (c). The inset in (b) shows a schematic representation of MoTe$_{2}$-EuO enclosed by a thermal source at temperature $T$. Horizontal dashed lines indicate two selected Fermi levels $E_{F1}=-1.17$ eV and $E_{F2}=0.2$ eV, where we obtain the MCE.}\label{fig1}
\end{figure}

\textit{Quantum-Thermodynamic model}.
To describe the low-energy spectrum and MCE of the MoTe$_{2}$-EuO heterostructure \cite{qi2015giant,Scharf2017},
we use a 3OTB model \cite{Liu2013} to include MEF effects. The model has relevant lattice symmetries, and has been proven to reliably describe TMDs on diverse situations, on magnetic substrates \cite{luo2019optical,cortes2019tunable,cortes2020reversible}, induced magnetic interactions, \cite{Avalos2016v2,avalosEdges2016}, and heterostructures \cite{Alsharari2018}.
The nearly commensuration of MoTe$_{2}$-EuO(111) \cite{qi2015giant,Zhang2016} (2.7\% lattice mismatch), incorporates the substrate effects into the pristine MoTe$_{2}$ as on-site magnetic exchange [see inset of Fig.\ \ref{fig1}(b)] and Rashba fields as \cite{cortes2019tunable}
\begin{equation}\label{MagnetizedHamiltonian}
\mathcal{H}_{\mathrm{MoTe}_{2}\mathrm{-EuO}}=\mathcal{H}_{\mathrm{MoTe}_{2}}+\mathcal{H}_{\mathrm{ex}}+\mathcal{H}_{R}. 
\end{equation}
$\mathcal{H}_{\text{MoTe}_{2}}$ is the pristine 2H phase TMD Hamiltonian \cite{Liu2013}, including intrinsic SOC and a matrix of hoppings considering next-nearest-neighbors, it is written in a basis of relevant transition metal $d$-orbitals, $\big\{ \left|d_{z^{2}},s_z \right\rangle$, $\left|d_{xy},s_z \right\rangle$, $\left|d_{x^{2}-y^{2}},s_z \right\rangle \big\}$, with spin $z$ component $s_z=\uparrow,\downarrow$ \cite{Liu2013}.
The induced MEF is spin diagonal, with blocks $\mathcal{H}_{\mathrm{ex},(\uparrow\uparrow)}=-\mathcal{H}_{\mathrm{ex},(\downarrow\downarrow)}=\mathrm{diag}\{-B_{c},-B_{v},-B_{v}\}$, where $B_{c}=206$ meV and $B_{v}=170$ meV correspond to conduction and valence exchange fields, respectively. Figure \ref{fig1}(a) shows the spin-polarized band structure from Eq.\ \ref{MagnetizedHamiltonian} for the suspended MoTe$_2$ 2D monolayer ($\mathcal{H}_{\mathrm{ex}}=\mathcal{H}_{R}=0$), and MoTe$_{2}$-EuO with $\mathcal{H}_{R}=0$ and different values of the exchange fields. The MEF breaks TRS, yielding large valley splittings in the valence and conduction bands \cite{qi2015giant,li2018large}, decreasing it as the MEF strength reduces. 
The Rashba Hamiltonian $\mathcal{H}_{R}$ in Eq.\ \ref{MagnetizedHamiltonian} is given by antidiagonal blocks,  mixing the spin and orbital components in the MoTe$_{2}$ monolayer with coupling $\lambda_R=72$ meV. All parameters are obtained from DFT calculations \cite{Liu2013,qi2015giant}. We first analyze the MCE for different values of the EuO MEFs and vanishing Rashba coupling ($\lambda_R=0$), then we include the Rashba effect for the MCE calculations, as we will see later.

In order to correlate both the 3OTB model and MCE, we numerically calculate the spin-polarized density of states (DOS), $D^{s_z}$, using a 2D Brillouin zone (BZ), i.e., $k_z=0$ in reciprocal $\pmb{k}$-space. We use a fine mesh of about ten million of $\pmb{k}$ points in the shaded area of the BZ [inset of Fig.\ \ref{fig1}(a)]. For every $\pmb{k}$-state we evaluate the eigenvalues from each band of the Hamiltonian of Eq.\ \ref{MagnetizedHamiltonian}. The DOS $D^{s_z}(E,B_{v},B_{c})$ depends on the eigenvalue with energy $E$, and both exchange fields $B_{v}$, $B_{c}$ induced on the MoTe$_{2}$ monolayer. Figures \ref{fig1}(b) and \ref{fig1}(c) show $D^{s_z}(E,B_{v},B_{c})$ for the suspended MoTe$_{2}$ monolayer and MoTe$_{2}$-EuO system. Note that for the suspended MoTe$_{2}$, the DOS are equivalent, i.e., $D^{\uparrow}(E,0,0)=D^{\downarrow}(E,0,0)$ (as expected from a nonmagnetic material), whereas $D^{\uparrow}$ ($D^{\downarrow}$) is downward (upward) shifted from the suspended MoTe$_{2}$ DOS. 

The total spin-polarized entropy $S_{tot}^{s_z}=S_{lat}(T)+S_{eM}^{s_z}(B_v,B_c,T)$ includes two terms, the entropy of the lattice $S_{lat}(T)$, giving account of the phonon contribution, where we assume it is only dependent on $T$; and the electromagnetic entropy $S_{eM}^{s_z}$ coming from the full Hamiltonian of Eq.\ \ref{MagnetizedHamiltonian}, and depending on the exchange fields and temperature. The electromagnetic entropy reads
\begin{equation}\label{entropy}
S_{eM}^{s_z}(B_{v},B_{c},T)=-k_{\text{B}} \int_{E_l}^{E_h}D^{s_z}(E,B_{v},B_{c})\mathcal{F}(n_\text{F})dE,
\end{equation}
with $E_{l(h)}$ the energy of the lowest (highest) occupied electronic eigenvalue. The probability of occupation of each eigenvalue is given by the Fermi-Dirac function distribution $n{_\text{F}}(E,T,\mu)=1/[e^{\beta(E-\mu)}+1]$ with $\beta=1/k_{\text{B}}T$, $k_{\text{B}}$ the Boltzmann constant, $\mu$ the chemical potential, and $T$ the heat-source temperature. In Eq.\ \ref{entropy} 
\begin{equation}\label{funcionnumero}
    \mathcal{F}(n_\text{F})=n_\text{F}\ln n_\text{F}+(1-n_\text{F})\ln(1-n_\text{F}), 
\end{equation}
is approximated by a Lorentzian-like function $L(E,T,\mu)=C/[e^{(|E-\mu|/2k_{\text{B}}T)^{3/2}}+1]$. By considering low and high $T$ values and $C=1.4$, we obtain excellent agreement between Eq.\ \ref{funcionnumero} and $L(E,T,\mu)$ with $-\mathcal{F}(n_\text{F})\approx L(E,T,\mu)$, so that Eq.\ \ref{entropy} transforms as \cite{cortes2021gate}
\begin{equation}\label{aproxentropy}
S_{eM}^{s_z}(B_{v},B_{c},T)\simeq k_{\text{B}} \int_{E_l}^{E_h}D^{s_z}(E,B_{v},B_{c})L(E,T,\mu)dE.
\end{equation} 
The main contribution of $L(E,T,\mu)$ to $S_{eM}^{s_z}$ is given by their temperature-dependent width, capturing more available states of the DOS as temperature increases \cite{cortes2021gate}. As reveals Eq.\ \ref{aproxentropy}, $S_{eM}^{s_z}$ is the link between the electronic and thermodynamic properties of the system. Therefore, the MCE in MoTe$_{2}$-EuO is obtained through constant temperature (isothermal) calculations of entropy changes $-\Delta S_{eM}^{s_z}$ between the entropy at zero MEF, and a final MEF as
\begin{equation}
\label{deltaS}
-\Delta S_{eM}^{s_z}
= S^{s_z}_{eM}(B_{v}=B_{c}=0,T)- S^{s_z}_{eM}(B_{v},B_{c},T),
\end{equation}
where we have ascribed the fact that $S_{lat}(T)$ is not affected by the MEFs, so that it does not present changes as $B_{v}$ and $B_{c}$ vary. In case to obtain $-\Delta S_{eM}^{s_z}>0$, we are in presence of the direct MCE, that is MoTe$_{2}$-EuO is capable to heat as $B_{v}$ and $B_{c}$ change. In the opposite case, when $-\Delta S_{eM}^{s_z}<0$, the system presents an inverse MCE and the sample cools down. This quantum-thermodynamic model provides an efficient and reliable approach to study spin-polarized properties of the proximity-induced MCE. 

\textit{Magnetocaloric reponse in MoTe$_{2}$-EuO}. In the electronic spectra of MoTe$_{2}$-EuO in Fig.\ \ref{fig1}, we have chosen two different Fermi levels to obtain the MCE as spin-polarized electromagnetic entropy changes $-\Delta S_{eM}^{s_z}$. These Fermi levels can be shifted by an overall gate field perpendicular to the TMD layer \cite{lazic2016effective,vzutic2019proximitized}, allowing for a tunable MCE in MoTe$_{2}$-EuO. The first Fermi level $\mu_1(T=0)=E_{F1}=-1.17$ eV is along the valence band, taking energies mainly from $K$ and $K'$ valleys, while $\mu_2(T=0)=E_{F2}=0.2$ eV crosses the $K-\Gamma-K'$ $\boldsymbol{k}$-path of the BZ, capturing additional valleys in the conduction band [see Fig.\ \ref{fig1}(a)]. As we use EuO in our calculations, it is important to know that their Curie temperature $T_C=69$ K \cite{ahn2005preparation,averyanov2019probing} can be larger when EuO is doped \cite{ott2006soft}, or when placed in close proximity to a 2D layer such as graphene \cite{averyanov2018high}. The latter shows that magnetic proximity effects including EuO are suitable for the study of magnetocaloric responses considering temperatures beyond $T_C$.

\begin{figure}[!h]
\centering
\includegraphics[width=\linewidth]{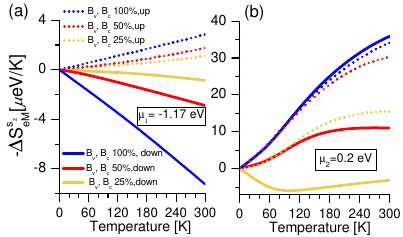}
\caption{Spin-polarized $-\Delta S_{eM}^{s_z}$ as a function of $T$ for MoTe$_{2}$-EuO when $\lambda_R=0$ and $B_v$ and $B_c$ are 25\% (yellow), 50\% (red) and 100\% (blue). The chemical potential is set to (a) $\mu_1=-1.17$ eV, and (b) $\mu_2=0.2$ eV. Note that $[\mu \text{eV/K}]\simeq [\text{J/kg K}]$ for the vertical axis, as the MoTe$_{2}$ unit cell in the 3OTB model has one Mo atom, and one mole of Mo weighs $\simeq 0.096$ Kg.}\label{fig2}
\end{figure}

Figure \ref{fig2} shows $-\Delta S_{eM}^{s_z}$ as a function of temperature for MoTe$_{2}$-EuO considering $\mu_1$ and $\mu_2$ and different values of the exchange fields $B_v$ and $B_c$. Clearly, the MCE has very different behavior for both Fermi levels. At $\mu_1=-1.17$ eV in Fig.\ \ref{fig2}(a), the MCE shows nearly a linear response as temperature increases, is strongly spin-polarized for all MEF values, and it is seen that weaker MEFs result in reduced MCE. At this Fermi level, the spin up polarized entropy changes are $-\Delta S^{\uparrow}_{eM} > 0$, contributing to heat the system, whereas for spin down $-\Delta S^{\downarrow}_{eM} < 0$, cooling down the sample up to room temperature. This dual behavior for the MCE occurs due to $D^{\uparrow}$ of MoTe$_2$ monolayer is larger than all other $D^{\uparrow}$ with MEFs different from zero, that is $D^{\uparrow}(\mu_1,B_{v}=0,B_{c}=0) > D{^\uparrow}(\mu_1,B_{v}\neq0,B_{c}\neq0)$, giving a positive MCE, while the opposite is true for $D^{\downarrow}$ at $\mu_1$.

In striking contrast, when the Fermi level $\mu=0.2$ eV is along the conduction band, as shown in Fig.\ \ref{fig2}(b), $-\Delta S_{eM}^{s_z}$ are nonlinearly spin-polarized and one order of magnitude larger than at $\mu_1$. For MEF of 100\% (blue lines), $-\Delta S_{eM}^{s_z}$ start to be spin polarized for $T> 120$ K, heating the sample for both spins. As $B_v$ and $B_c$ decrease to 50\%, both spin-polarized components of $-\Delta S_{eM}^{s_z}$ (red lines) still heats up, and $-\Delta S^{\uparrow}_{eM}$ (dotted red line) has the same value as the entropy changes for 100\% of EuO up to $T\approx 90$ K because the $L$ function (Eq.\ \ref{aproxentropy}) captures the same amount of states of the DOS. $-\Delta S_{eM}^{s_z}$ for MEF of 25\% are fully spin polarized from 0 to 300 K, $-\Delta S_{eM}^{\uparrow}$ (yellow dotted line) heats up, while $-\Delta S_{eM}^{\downarrow}$ (yellow solid line) cools down. This behavior is fully related to the magnitudes of $D^{s_z}$ as we describe below.

\begin{figure}[!h]
\centering
\includegraphics[width=\linewidth]{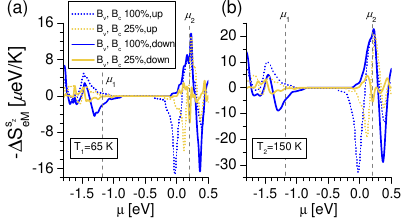}
\caption{Spin-polarized $-\Delta S_{eM}^{s_z}$ as a function of $\mu$ for MoTe$_{2}$-EuO when $\lambda_R=0$ and $B_v$, $B_c$ are 25\% (yellow) and 100\% (blue). The temperature is set to (a) $T_1=65$ K below the $T_C$ of EuO, and (b) $T=150$ K above the $T_C$ of EuO. Dashed vertical lines indicate the chemical potentials used in Fig.\ \ref{fig2}.}\label{fig3}
\end{figure}

We calculated $-\Delta S_{eM}^{s_z}$ as a function of the chemical potential for 100\% and 25\% of EuO, and selecting two values of temperature (below and above the $T_C$ of EuO). Figure\ \ref{fig3}(a) shows $-\Delta S_{eM}^{s_z}$ at $T=65$ K, and Fig.\ \ref{fig3}(b) at $T=150$ K. Because the MEFs compete with the intrinsic SOC of the 2D MoTe$_2$ monolayer, $-\Delta S_{eM}^{s_z}$ is strongly spin-polarized as a function of $\mu$ for both values of $T$ and MEF strengths of EuO. For both temperatures, and in the range of energies of the valence valley polarization ($-1.5\ \text{eV} \lesssim \mu \lesssim -0.9\ \text{eV}$), $-\Delta S_{eM}^{s_z}$ for full EuO MEF (blue lines) show positive (heating) and negative (cooling) peaks near where $D^{s_z}$ present maxima, i.e., at $\mu \simeq -1.5$ eV for suspended MoTe$_2$, and at $\mu \simeq -1.3$ eV for $D^{\downarrow}$ of 100\% EuO respectively, see Figs.\ \ref{fig1}(b) and \ref{fig1}(c). $-\Delta S_{eM}^{s_z}$ vanishes in the band gap region of the combined MoTe$_2$-EuO system ($-0.9\ \text{eV} \lesssim \mu \lesssim -0.4\ \text{eV}$) for both values of $T$ and MEFs strengths, as there are not available states in the energy spectra. 

For Fermi levels in the conduction bands ($-0.4\ \text{eV} \lesssim \mu \lesssim 0.5\ \text{eV}$), there are large positive and negative peaks for $-\Delta S_{eM}^{s_z}$. For 100\% of EuO, a clear negative peak is near $\mu \simeq 0$ due to the maxima of $D^{\uparrow}$ at this Fermi level [Fig.\ \ref{fig1}(c)]. Two other positive peaks are almost non spin-polarized about $\mu_2= 0.2$ eV [in agreement with Fig.\ \ref{fig3}(b)] as $D^{s_z}$ of the suspended MoTe$_2$ monolayer presents maxima near this energy. Two negative peaks are about $\mu \simeq 0.4$ eV, where $D^{\downarrow}$ shows a large peak, while $D^{\uparrow}$ is reduced, giving different $-\Delta S_{eM}^{s_z}$ values near this Fermi level for 100\% of EuO. Notice that $-\Delta S_{eM}^{s_z}$ is about two times larger for $T=150$ K because the $L$ function (Eq.\ \ref{aproxentropy}) captures more states of the DOS when temperature increases. As the MEFs decrease to 25\% of EuO, some $-\Delta S_{eM}^{s_z}$ conduction peaks are shifted from 100\% of EuO, and have opposite spin-polarization at $\mu_2=0.2$ eV due to similar $D^\downarrow$ magnitudes for MoTe$_2$ with 25\% of EuO [Fig.\ \ref{fig1}(b)].

\begin{figure}[!h]
\centering
\includegraphics[width=\linewidth]{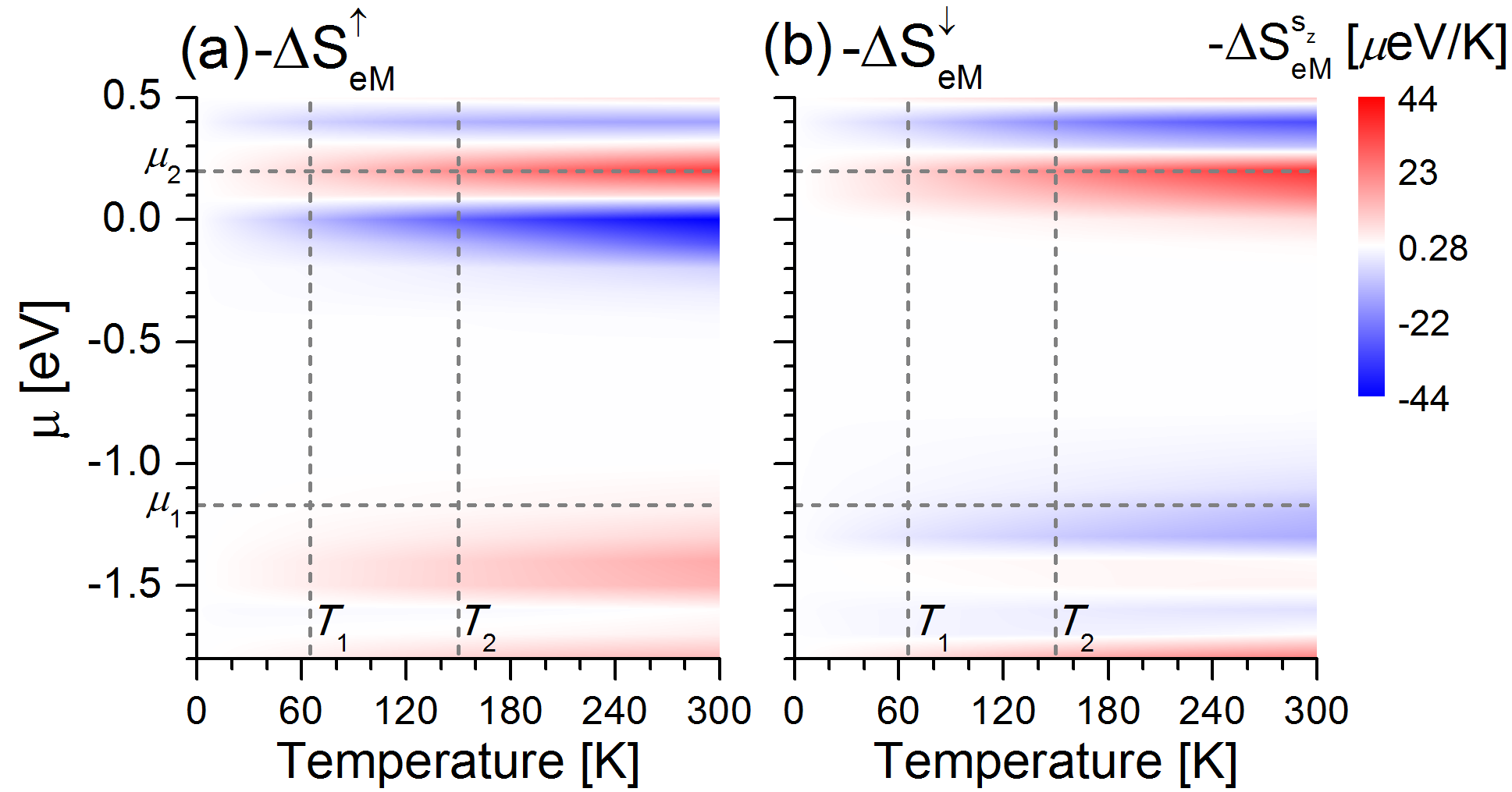}
\caption{Contour plots for $-\Delta S_{eM}^{s_z}$ as a function of the chemical potential and temperature for MoTe$_{2}$-EuO when $\lambda_R=0$ and $B_v$, $B_c$ are 100\%, (a) spin up, (b) spin down. Horizontal gray dashed lines indicate the two different $\mu$ values for $-\Delta S_{eM}^{s_z}$ calculations in Fig.\ \ref{fig2}. Vertical gray dashed lines show temperatures used in Fig.\ \ref{fig3}, $T_1=65$ K, $T_2=150$ K. Color bar indicates positive (negative) entropy changes as red (blue) gradient.}\label{fig4}
\end{figure}

Because the entropy changes for the MoTe$_2$-EuO heterostructure are highly dependent on the chemical potential and temperature, we present contour plots for the spin polarized $-\Delta S_{eM}^{s_z}$ as a function of $\mu$ and $T$ for EuO full MEF, $-\Delta S^{\uparrow}_{eM}$ in Fig.\ \ref{fig4}(a), and $-\Delta S^{\downarrow}_{eM}$ Fig.\ \ref{fig4}(b). We can see that into the valence valley polarization energy zone $\sim$ -1.3 to -1 eV, the system is capable to heats for spin up, and cools for spin down. As higher Fermi levels are reached in the conduction bands ($\mu \gtrsim -0.2$ eV), $-\Delta S^{\uparrow}_{eM}$ cools, then heats, and again cools, while $-\Delta S^{\downarrow}_{eM}$ heats and then cools. Accordingly, one could modulate the spin-polarized MCE in the MoTe$_2$-EuO system to cooling or simultaneously heating, by tuning the Fermi level across the structure.

\begin{figure}[!h]
\centering
\includegraphics[width=\linewidth]{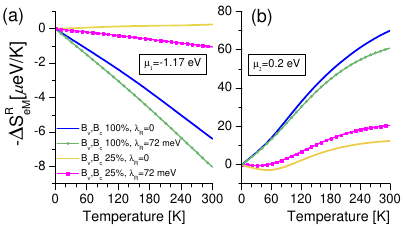}
\caption{$-\Delta S_{eM}^{R}$ as a function of $T$ for (a) $\mu_{1}=-1.17$ eV and (b) $\mu_{2}=0.2$ eV for MoTe$_{2}$-EuO when $\lambda_R=0$ (solid lines) and $\lambda_R=72$ meV (symbol lines), $B_v$ and $B_c$ are 25\% (yellow and pink) and 100\% (green and blue) of EuO.}\label{fig5}
\end{figure}

The broken spatial symmetry generated by the proximity with the EuO substrate generates an interfacial Rashba field \cite{kane2005quantum,ochoa2013spin,frank2018,cortes2019tunable,cortes2020reversible} that produces in-plane spin contributions $s_x$, $s_y$ on the MoTe$_2$ 2D monolayer \cite{qi2015giant}. We have incorporated the Rashba field in our calculations for exchange fields of 100\% and 25\%, as shown in Fig.\ \ref{fig5} for $-\Delta S_{eM}^{R}$ as a function of $T$. In order to compare the effect of the Rashba field, we have used the previous MCE results when $\lambda_R=0$ (Fig.\ \ref{fig2}), through $-\Delta S_{eM}^{\uparrow+\downarrow}=-\Delta S_{eM}^{\uparrow}+(-\Delta S_{eM}^{\downarrow})$. 

Figure \ref{fig5}(a) shows that at $\mu_{1}=-1.17$ eV, the Rashba field linearly enhances the entropy changes for EuO full MEF (symbol green line), and producing a negative MCE (cooling) for both EuO MEF strengths. At $\mu_{2}=0.2$ eV in Fig.\ \ref{fig5}(b), the entropy changes for 100\% of EuO have the same values with and without the Rashba field up to near the $T_C$ of EuO. As temperature increases, the Rashba field does not improve the entropy changes up to room temperature. A contrasting behavior is seen for MEFs of 25\% of EuO, in which the entropy changes present different values for all temperatures, and the Rashba field enhanced it starting in $T\approx 100$ K. These results suggest that for full EuO MEFs, the in-plane spin components are stronger in the valence band than in the conduction band as temperature increases, generating an enhanced MCE when the Rashba field is present.   

\textit{Conclusions}. 
Time reversal symmetry breaking on the 2D MoTe$_2$ monolayer because of induced EuO magnetic exchange fields result in density of states variations, producing large spin-polarized entropy changes across the valley splitting energy-momentum space of the combined MoTe$_2$-EuO system. By gating the heterostructure, tunable spin-dependent heating and/or cooling can be achieved in the valence and conduction energy zones, improving it when the Rashba field is taken into account in the valence band. The proximity-induced magnetocaloric effect in semiconducting MoTe$_2$ reveals spin-dependent quantum-thermodynamic responses that could be used in the design of novel atomistic cooling technologies. 

\textit{Acknowledgments}.
N.C. acknowledges support from ANID Fondecyt Postdoctoral Grant No. 3200658, F.J.P. acknowledges support from ANID Fondecyt, Iniciaci\'on en Investigaci\'on 2020 Grant No. 11200032, and the financial support of USM-DGIIE. O.N. and P.V. acknowledge support from ANID PIA/Basal AFB18000, and P.V. acknowledges support from ANID Fondecyt Grant No. 1210312.

\bibliography{bib}
\email{natalia.cortesm@usm.cl}

\end{document}